\documentclass{article}
\usepackage{amsfonts}
\usepackage{amssymb,amsmath}
\usepackage{graphicx}
\usepackage{dcolumn}
\usepackage{bm}
\usepackage{color}
\usepackage{color}
\usepackage{amsmath}
\def\be{\begin{equation}}
\def\ee{\end{equation}}
\def\arr{\begin{array}{rll}}
\def\ea{\end{array}}
\def\bea{\begin{eqnarray}}
\def\eea{\end{eqnarray}}

\def\N2{$N{=}2$}

\def\>{\rangle}
\def\<{\langle}
\def\+{\dagger}
\def\={\ =\ }

\begin{document}
\renewcommand{\thefootnote}{\fnsymbol{footnote}}
\setcounter{page}{1}
\thispagestyle{empty}
\begin{center}
{\Large\bf Landau problem on the ellipsoid, hyperboloid and paraboloid of revolution}\\[7mm]

{\large Eva Gevorgyan$^a$, Armen Nersessian$^{a,b}$, Vadim
Ohanyan$^{a,c,d}$, Evgeny Tolkachev$^{e}$ }
\\[3mm]
 $\;^a${\sl Yerevan State University, 1 Alex
Manoogian St., Yerevan, 0025, Armenia}\\
$\;^b${\sl Tomsk Polytechnic University, Lenin Ave. 30, 634050 Tomsk, Russia}\\
 $\;^b${\sl Institut f\"{u}r Theoretische Physik, Leibniz Universit\"{a}t Hannover, 30167 Hannover, Germany}\\
$\;^c${\sl ICTP, Strada Costiera 11, I-34151 Trieste, Italy}\\
 $\;^c$ {\sl Institute of
Physics, Belarus National Academy of Sciences, F.Skaryna Av.,
68, Minsk, 220072, Belarus }
\end{center}

\begin{abstract}
We define the Landau problem on two-dimensional
ellipsoid, hyperboloid and paraboloid of revolution. Starting form the two-center
MICZ-Kepler system and making the reduction into these
two-dimensional surfaces we obtain the Hamiltonians of the charged
particle moving on the corresponding surface of
 revolution in the magnetic field conserving the symmetry of the two-dimensional surface (Landau problem). For each
  case we figure out the values of parameter for which the qualitative character of the motion coincides with that of a
   free particle moving on the save two-dimensional surface. For the case of finite trajectories we construct the action-angle variables.
\end{abstract}


\section{Introduction }
The classical Landau problem refers to a problem of the particle
moving in the plane in a homogeneous constant magnetic field
orthogonal to the plane\cite{ll}. From the symmetry point of view
the Landau problem is characterized by the presence of a magnetic
field which preserves the symmetry of the free particle system.
There is a natural generalization of the Landau problem for
two-dimensional sphere which has been used for constructing the
translationally invariant version of Laughlin's states for
fractional quantum Hall effect\cite{la,haldane}. The Landau problem
on the $S^2$ is defined as
 a problem of particle motion on the surface of the two-dimensional sphere with a Dirac monopole placed in its center.
 Thus, the corresponding Hamiltonian is rotationally symmetric. Generally speaking, to generalize the Landau problem
 for the case of a particle moving on a two-dimensional manifold $\mathcal{M}$ one should find a way to include a magnetic
  field which preserves the isometries of $\mathcal{M}$, i.e. $T_2\bigotimes SO(2)$ in case of $\mathbb{R}^2$ and $SO(3)$ in case
   of $S^2$ etc. For the case of higher-dimensional manifolds the situation is more complicated and, so far, only the Landau
    problems on the so-called Hopf's spheres ($S^2$, $S^4$ and $S^8$) have been understood in details, because the structure
    of the corresponding magnetic monopole field can be easily obtained within the Hamiltonian reduction corresponding to the
     Hopf's fibrations $S^3/S^1\simeq S^2$, $S^7/S^3\simeq S^4$ and $S^{15}/S^7\simeq S^8$.
     For instance,
      four-dimensional or quaternionic Landau problem is defined as a  $SO(5)$-invariant system of
      isospin particle on $S^4$ interacting with  BPST instanton (equivalently, from the viewpoint of
       ambient five-dimensional Euclidean space, particle on four-sphere interacting with the field of
       $SU(2)$ Yang monopole located at the center of the sphere)\cite{science}. The higher dimensional analogs of these
       configurations are defined, respectively, on the complex projective spaces $\mathbb{CP}^n$ \cite{nair} and
       quaternionic projective spaces $\mathbb{HP}^n$ \cite{bkny}. It is rather straightforward to construct their non-compact analogs \cite{ahmed}, as well as  further
       generalization of these  systems for the corresponding (complex and quaternionic) Grassmanians. Let us
    also mention  "octonionic Landau problem" on eight-dimensional
    sphere considered in \cite{octonions}.

      From the quantum mechanics point of view specific feature of
      the Landau problem is the similarity between its
      spectrum and that of a free particle.
      Thus, the inclusion of the external gauge field (i.e. in transition
      from the free particle to Landau problem) yields the exclusion of the lower part of the energy spectrum,
      resulting in the degeneracy of the ground state. Just thanks to these degeneracy Landau problem plays
      such an important role in many areas of theoretical physics and, particularly forms the background
      of the various models of Hall effect.

    The interesting and important question is the construction of the analog of Landau problem on non-symmetrical spaces.
    In the present paper we try to answer this question
    for the particular case of the simplest two-dimensional non-symmetric spaces,
    which are  the  ellipsoid, the hyperboloid and the paraboloid of revolution.

    For this purpose we use the following  analogy. It is easy to check, that placing the Dirac monopole in the center
    of $SO(3)$-symmetric three-dimensional system supplied by the adding of specific "centrifugal" potential term one can preserve qualitative
     properties of the
    initial system. The only change in its spectrum is the modification of the range of possible values of the orbital quantum number.
    In classical mechanics introduction of the monopole leads to the shift of the orbital  plane to the fixed angle (see, e.g.
    \cite{lnp}).
    The  Landau problem on $S^2$ can be considered as a spherical part of this three-dimensional system with a Dirac monopole obtained by the fixing of the radial
    coordinate (and, respectively, setting to zero its conjugated momentum).  It can also be viewed  as a spherical part  of the so-called MICZ-Kepler
    system \cite{micz} describing the
     motion of charged particle in the field of Dirac dyon
     (scalar particle carrying both electric and magnetic charges):
     potential term of this system  having dependence  on the radius only, does not have any
      impact on the spherical part of Hamiltonian. Planar Landau problem (on $\mathbb{R}^2$) can be obtained from that
       on $S^2$ by taking the limit of infinite radius of the sphere.

    The $SO(5)$-symmetric five-dimensional system behaves in completely similar way upon
     placing of the Yang monopole in its center and adding the "centrifugal"
     term. This is the case, particularly, for the five-dimensional analog of MICZ-Kepler system known also as
     $SU(2)$-Kepler" system\cite{tea,mar,ple,ple2}.
      Respectively,  the (quaternionic) Landau problem on $S^4$ can be extracted from Yang monopole
    in the similar manner, i.e. by fixing the radius of the ambient system \cite{mny}.

Few years ago  the two-center MICZ-Kepler systems were
constructed, where the inclusion of monopoles yields, qualitatively,
the same changes as in one-monopole  MICZ-Kepler--like systems
\cite{KNO,NO} (see also \cite{BO,denef}). The main statement of
those works is the following: let we have some one-particle
three-dimensional system admitting separation of variables in elliptic
coordinates. Placing the Dirac monopoles in the foci of the elliptic
coordinate system and simultaneous adding the analog of
"centrifugal" term preserve separability of variables. In the
limiting case, when one of the foci is placed at infinity (in this
case the corresponding electric and magnetic fields of the monopole
transform to the constant uniform ones, parallel to each other ),
 we get the similar statement concerning the parabolic coordinates.

With the aid of this system we construct the analog of (classical) Landau
problem on the  ellipsoid and  hyperboloid of revolution.
 We  start with the two-center MICZ-Kepler system and restrict
it to the ellipsoid or the hyperboloid by fixing the value of the
corresponding elliptic coordinate (and setting its conjugated
momentum to zero). The corresponding magnetic field of the pair of monopoles reduces to a
certain configuration preserving the symmetry of the underlying two-dimensional surface.
Obviously, the system obtained in this way is the generalization of the Landau problem
for the two-dimensional second order compact surfaces with axial symmetry (ellipsoid and hyperboloid).
Considering the classical solution within the Hamilton-Jacobi formalism, we then can specify the values of the parameters of the initial system
at which the emergent two-dimensional Landau problems are solved in terms of the same elliptic integrals as the problem of the free particle moving on the same surfaces.
Thus, at these specific values of the parameters the Landau problem has qualitatively the same trajectories as the free system. This particular situation can be referred to as
"the reducible Landau problem``.
 This results in the restriction on the electric and magnetic
charges of the background dyons. They are specified by the following values of
electric and magnetic charges
   \begin{itemize}
\item Ellipsoid: $e_{1}=e_{2}, g_{1}=-g_{2}$,
\end{itemize}
i.e. we get, on the background, magnetic dipole with the equal electric charges on the poles.
So, in the case of the ellipsoid the reducible Landau problem corresponds to the physically relevant configuration.
  \begin{itemize}
\item Hyperboloid: $e_{1}=-e_{2}, g_{1}=g_{2}$,
\end{itemize}
i.e. the background configuration is the electric dipole with an equally magnetically charged poles. \\

In the limiting, parabolic case we get less interesting background configuration
 \begin{itemize}
\item Paraboloid: dyon in the focus with arbitrary electric and magnetic charge,
vanishing constant uniform electric  field parallel to the axes of paraboloid ($B=0, {\cal E}= 0$).
\end{itemize}
Since classical solution of these generalized Landau problems are given by elliptic integrals,
corresponding quantum-mechanical systems are not exactly solvable. Nevertheless, it is interesting to check, at least by numerical simulations,
wether they possess the degenerate ground state \footnote{ In the remarkable paper of Ferapontov and Veselov \cite{fv},
 factorization of two-dimensional Schr\"odinger operator in the presence of magnetic field has been investigated.
Particularly, they presented there quantum-mechanical system on ellipsoid,  admitting factorization of the Hamiltonian,
and constructed their ground states. In fact, these systems could also called "elliptic Landau problem(s)".
 However, they are different from the one proposed by us. Thus, quantum counterpart of our Hamiltonians could not be factorizable ones.}.
 Let us notice, that constructed systems are so simple and visible, that they look as exercises
  from the textbook on electrodynamics or  classical mechanics. However we could not find them in  the existing textbooks. Probably, they were
  presented in some journal or textbook in XIX or XVIII century.
  In any case, we assume, that presentation of these results  is useful, at least, from pedagogical point of view.

  The paper is organized as follows.
 In {\sl Section 2} we  present the Hamilton-Jacobi treatment of the free particle on the ellipsoid, hyperboloid and paraboloid of revolution. In {\sl Section 3}
 we consider the restriction of the two-center MICZ-Kepler systems to ellipsoid, hyperboloid and paraboloid and find the conditions of the similarity of
 the classical motion to the case of a free particle. The last
 {\sl Section} contains the conclusion and discussion.

  \section{Free particle on the ellipsoid, hyperboloid and paraboloid of revolution}
  Let us start with a Hamiltonian of a free particle moving in a tree-dimensional space written down in standard elliptic coordinates:
  \begin{eqnarray}\label{Ham1}
  \mathcal{H}=\frac{1}{2a^2\left(\xi^2-\eta^2
  \right)}\left\{(\xi^2-1)p_\xi^2+(1-\eta^2)p_\eta^2+\frac{\xi^2-\eta^2}{(\xi^2-1)(1-\eta^2)}p_{\varphi}^2\right\},
  \end{eqnarray}
  where the  elliptic coordinates $\xi, \eta$ and $\varphi$
  are related to the standard cylindrical ones ($\rho, z, \varphi$) via the following relations:
  \begin{eqnarray}\label{el_co_1}
  \rho=a \sqrt{(\xi^2-1)(1-\eta^2)},\quad z=a\xi\eta, \quad \varphi=\varphi,
  \end{eqnarray}
  where $a$ is a parameter whose meaning become more clear if one introduces two "origins"
   in $\mathbb{R}^3$ situated on the $z$-axis and separated by the distance $2a$, then $r_1$ and $r_2$ are
    the distances between arbitrary point $P$ of the $\mathbb{R}^3$ and two "origins" with coordinates
  $(0,0,-a)$ and $(0,0,a)$ respectively.
  It is easy yo see that
  \begin{eqnarray}
 \xi=\frac{r_1+r_2}{2a}, \quad \eta=\frac{r_1-r_2}{2a}, \label{el_co_2}
  \end{eqnarray}
  where $\xi\in\left[1,\infty\right)$ and
  $\eta\in\left[-1,1\right]$.
  Obviously, the coordinate surfaces are ellipsoids with the foci at  $(0,0,-a)$ and $(0,0,a)$ at fixed values of $\xi$:
  \begin{eqnarray}\label{ellipsoids}
  \frac{z^2}{a^2 \xi^2}+\frac{\rho^2}{a^2 (\xi^2-1)}=1,
  \end{eqnarray}
  and hyperboloids,
  \begin{eqnarray}\label{hyperboloids}
   \frac{z^2}{a^2 \eta^2}-\frac{\rho^2}{a^2 (1-\eta^2)}=1
  \end{eqnarray}
  at fixed values of $\eta$.

  \subsection*{Ellipsoid}
  Let us now consider the particle moving on the ellipsoid obtained by rotation of the ellipse around $z$-axis .
   The corresponding Hamiltonian can be obtained from Eq. (\ref{Ham1})
  by setting $\xi=1/e=const $ and $p_{\xi}=0$. The constant $e$ has
  the obvious meaning of the eccentricity of the ellipse which is
  any  section of the ellipsoid $\xi=1/e=const$ with the plane
  containing $z$-axis. Thus, the corresponding Hamiltonian is
  \begin{eqnarray}\label{ham_elip_e}
    \mathcal{H}_{el}=\frac{e^2}{2a^2\left(1-e^2\eta^2
  \right)}\left\{(1-\eta^2)p_\eta^2+\frac{1-e^2\eta^2}{(1-e^2)(1-\eta^2)}p_{\varphi}^2
  \right\},
  \end{eqnarray}
 Let us use the Hamilton-Jacobi method to solve the problem of motion of the free particle on ellipsoid.
 One should look for the generating function (the action) in the following form
  \begin{eqnarray}
  S_{el}=-Et+p_{\varphi}\varphi+S_{\eta}, \label{gen_fun}
  \end{eqnarray}
  where $E$ is the energy of the system and
\be\label{S_el_eta}
S_{\eta}=\int p_{\eta}d \eta = \frac{1}{e\sqrt{1-e^2}}\int \sqrt{2 a^2(1-e^2)
(1-e^2\eta^2)(1-\eta^2)E-e^2(1-e^2\eta^2)p_{\varphi}^2}\;\frac{d\eta}{1-\eta^2}
\ee
According to standard technique, the equations
of motion in parametric form are obtained by taking derivatives of
the generating function with respect to constants $E$ and
$p_{\varphi}$ and setting the result to zero:
\be
t=\frac{\partial S_\eta}{\partial E},\qquad \varphi=-\frac{\partial S_\eta}{\partial p_\varphi}.
\ee
%
For our purposes it is convenient to make the
 change of variable $1-\eta^{2}=x$
($x\in\left[0,1\right]$), and represent $S_\eta$ as follows
\be\label{S_el_eta_x}
S_{\eta}=-a\sqrt{\frac{E}{2}}\int\sqrt{\frac{x^2+b_{1}x+b_2}{1-x}}\frac{dx}{|x|}
\ee
where
\be\label{b1b2}
b_{1}=\frac{2a^2E(1-e^2)^2-p_{\varphi}^{2}e^4}{2a^2e^2E(1-e^2)},\quad b_{2}=-\frac{p_{\varphi}^{2}}{2a^{2}E}.
\ee

 \subsection*{Hyperboloid}
 The Hamiltonian (\ref{Ham1}) also contains another limiting case, the case of particle
 moving on the (two-sheet) hyperboloid. In order to obtain that one should set
    $\eta=1/e=const$ and  $p_{\eta}=0$. Then, in full analogy with the previous case the constant
    value of $\eta$ is in the same relationship to the eccentricity of the corresponding hyperbola $e$.
    Thus, the Hamiltonian of the free particle moving on the two-sheet hyperboloid is
\begin{eqnarray}
   \mathcal{H}_{hyp}=\frac{e^2}{2a^2\left(1-e^2\xi^2
  \right)}\left\{(1-\xi^2)p_\xi^2+\frac{1-e^2\xi^2}{(1-e^2)(1-\xi^2)}p_{\varphi}^2
  \right\}, \label{ham_hyp_e}
\end{eqnarray}
Then, within the standard Hamilton-Jacobi formalism, one obtains the generating function
\begin{eqnarray}
 S_{hyp}=-Et+p_{\varphi}\varphi+S_{\xi}, \label{gen_fun_h}
\end{eqnarray}
where
\begin{eqnarray}
S_{\xi}=\frac{1}{e\sqrt{1-e^2}}\int \sqrt{{2 a^2
(1-e^2)(1-e^2\xi^2)(1-\xi^2) E- e^2(1-e^2\xi^2)p_{\varphi}^2}}\;\frac{d\xi}{1-\xi^2}.
\label{sh}\end{eqnarray}
The equations of motion read $t=\partial S_\xi /\partial E$, $\varphi=\partial S_\xi /\partial p_\varphi$.
Making the suitable variables change
$1-\xi^{2}=x$ ($x\in\left(-\infty,0\right]$) one will have for
$S_{\xi}$ the same integral expression as in the case of ellipsoid
from Eq.(\ref{S_el_eta_x}) with the same values of the constants from
Eq.(\ref{b1b2}). Thus, the motion of the free particle on  both ellipsoid and hyperboloid is given by the same
expression for generating function. The only difference is
the range of the variable $x$.
\subsection*{Paraboloid}
Now, let us consider the Hamiltonian of the particle moving in the three-dimensional Euclidian space
 written down in the parabolic coordinates:
  \begin{eqnarray}\label{Ham_p}
  \mathcal{H}_{par}=\frac{1}{2(\xi+\eta)}\left(4\xi p_{\xi}^{2}
  +4\eta p_{\eta}^{2}+\frac{\xi+\eta}{\eta\xi}p_{\varphi}^{2}\right)
 ,
  \end{eqnarray}
 where 
   $\xi$, $\eta$ and $\varphi$ are expressed in standard cylindrical coordinates($\rho$, $z$,
   and $\varphi$) in the following way:
  \begin{eqnarray}
  \rho=\sqrt{\xi\eta},\quad z=\frac{\xi-\eta}{2}, \quad \varphi=\varphi, \label{el_co_1}
  \end{eqnarray}
  where both $\xi$ and $\eta$ take values within
  $\left[0,\infty\right)$. It is easy to see that the surfaces of
  constant $\xi$ and $\eta$ are two families of concentric
  paraboloids of revolution with the symmetry axis pointed along $z$:
  \be\label{parabaloids}
\frac{\rho^{2}}{\xi}-\xi+2z=0,\;\;\;\frac{\rho^2}{\eta}+\eta-2z=0.
  \ee
 Let us now consider the motion of the free particle on the
 paraboloid which can be obtained form the Hamiltonian (\ref{Ham_p}) by assuming
$\eta=p/2=const$, with  $p$ being the
parameter of the corresponding parabola which is any section of the
paraboloid with the plane containing $z$-axis,
\be
\mathcal {H}_{par}=\frac{1}{p+2\xi}
\left(4\xi p_{\xi}^{2}+\frac{p+2\xi}{p\xi}p_{\varphi}^{2}\right)
\ee
In the Hamilton-Jacobi formalism the corresponding generating
function is
\be S_{par}=-Et+p_{\varphi}\varphi+ S_\xi,\qquad S_\xi=\frac{1}{2\sqrt{p}}\int
\sqrt{(pE\xi
 -p_{\varphi}^{2})(p+2\xi)}\;\frac{d\xi}{\xi} =\sqrt{\frac{E}{2}}\int\sqrt{\xi^2+b_{1}\xi+b_{2}}\;\frac{d\xi}{\xi}
 \label{S_par_xi}\ee
where
 \be
 b_{1}=\frac{p}{2}-
 \frac{p_{\varphi}^{2}}{pE}, \quad b_{2}=-\frac{p_{\varphi}^{2}}{2E}
 \ee
 The equations of motion are defined by the expressions $t=\partial S_\xi/\partial E $, and $\varphi=\partial S_\xi/\partial p_\varphi $.
The expressions for  generating functions of the free particle on
   the ellipsoid, hyperboloid and paraboloid of revolution are the
  reference expressions for the formulation of the Landau problems on these surfaces given in the next section.

\section{Two-center MICZ-Kepler system  and the Landau problem on ellipsoid, hyperboloid and paraboloid}
 For the construction of the Landau problem on ellipsoid and hyperboloid we use
 the so-called two-center MICZ-Kepler system\cite{KNO,NO}.
 The system describes the charged particle with {\sl unit} electric charge moving in the electric and magnetic
 fields of two dyons (electrically charged Dirac monopoles) with electric (magnetic) charges $q_{1,2}$ ($g_{1,2}$)
 which are fixed at the points with coordinates   $(0,0,-a)$ and $(0,0,a)$ respectively.
  Choosing the appropriate gauge for the vector-potential of the Dirac monopoles
\be\label{pot}
 A_r^{1,2}=A_\theta^{1,2}=0, \quad A_\varphi^{1,2}=g_{1,2}\cos
 \theta_{1,2},
 \ee
one arrives at the following Hamiltonian \cite{KNO}
\be
 \mathcal{H}=\frac{p_r^2}{2}+\frac{p_\theta^2}{2r^2}+
 \frac{\left(p_\varphi-g_1 \cos \theta_1-g_2 \cos\theta_2 \right)^2}{2 r^2 \sin^2 \theta}
 +\frac{1}{2}\left( \frac{g_1}{r_1}+\frac{g_2}{r_2}
 \right)^2+\frac{q_1}{r_1}+\frac{q_2}{r_2}. \label{ham_two}
\ee
where   $r_{1,2}$ stand for
the distance between the left(right) dyon and the moving particle. In the Ref.\cite{KNO}
it is shown that this Hamiltonian admits
separation of variables in elliptic coordinates. Being presented in
these coordinates, it has the following form:
\be\label{Ham_MICZ}
  \mathcal{H}=\frac{1}{2a^2\left(\xi^2-\eta^2
  \right)}\left\{(\xi^2-1)p_\xi^2+(1-\eta^2)p_\eta^2+\frac{\xi^2-\eta^2}{(\xi^2-1)(1-\eta^2)}p_{\varphi}^2+V(\xi)+W(\eta)\right\}
\ee
 where
 \be
 V(\xi)=\frac{g_{+}^{2}-2p_{\varphi}g_{-}\xi}{\xi^{2}-1}+2aq_{+}\xi, \qquad
 W(\eta)=\frac{g_{-}^{2}-2p_{\varphi}g_{+}\eta}{1-\eta^{2}}+2aq_{-}\eta
 \ee
and the following notations are adopted: $g_{\pm}=g_1 \pm g_2$ and
$q_{\pm}=q_1 \pm q_2$.
It is clear, that fixing the values of coordinates $\eta$ or $\xi$, we shall get
  the systems on ellipsoid or hyperboloid of revolution, respectively.
As mentioned above, the Hamiltonian of a
charged particle moving in the superposition of the electric and
magnetic fields of fixed dyon (with electric charge $q$ and magnetic charge $g$) and  parallel  constant uniform electric and
magnetic fields ${\cal E} $ and $B$ can be obtained from the two-center MICZ-Kepler
Hamiltonian by placing one of the dyons at infinity \cite{KNO,NO,BO}.
 This limiting Hamiltonian takes the following form:
 \be
 \mathcal{H}=\frac{p_r^2}{2}+\frac{p_\theta^2}{2r^2}+
 \frac{\left(p_\varphi-g\cos \theta-\frac{1}{2}Br^{2}\sin^2 \theta \right)^2}{2 r^2 \sin^2 \theta}
 +\frac{1}{2}\left( \frac{g}{r}+ B z
 \right)^2+\frac{q}{r}-{\cal E} z. \label{ham_two}
 \ee
 It admits the separation of variables in
 parabolic coordinates. In these coordinates it reads
\be\label{Ham_par_ZS}
 {\mathcal{H}}=\frac{ 1}{2(\xi+\eta)}\left( 4\xi p_\xi^2+4\eta
p_\eta^2 +V(\xi)+W(\eta)\right)
-\frac{1}{2}B p_\varphi ,
\ee
where
\be\label{par_pot}
{{V}}(\xi)=\frac{(p_\varphi+g)^2}{\xi}+3gB\xi-{\cal E}\xi^2+\frac{B^2 }{4}\xi^3+2q, \qquad
{{W}}(\eta)=\frac{(p_\varphi-g)^2}{\eta}-3gB\eta+{\cal E}\eta^2+\frac{B^2}{4}\eta^3+2q.
\ee
It is clear, that fixing the coordinate $\xi$ we will get system on the  paraboloid of revolution.

\subsection*{The Landau problem on ellipsoid}
 Let us now set $\xi={1}/{e}=const$ and $p_{\xi}=0$, $0<e<1$, which corresponds to the projection of the three-dimensional system to the one of
 two-dimensional  ellipsoids of revolution. Then, the Hamiltonian (\ref{Ham_MICZ})  reduces to the Hamiltonian
 of the  particle (with {\sl unit} electric charge) on the ellipsoid of revolution with two dyons placed in its foci,
 with arbitrary values of  electric and magnetic charges, $q_{1,2}$
and $g_{1,2}$:
\be\label{el_MICZ}
  \mathcal{H}_{el}=\frac{e^2}{2a^2\left(1-e^2\eta^2
  \right)}\left\{(1-\eta^2)p_\eta^2+\frac{1-e^2\eta^2}{(1-e^2)(1-\eta^2)}p_{\varphi}^2
  +\frac{g_{-}^{2}-2p_{\varphi}g_{+}\eta}{1-\eta^{2}}+2aq_{-}\eta+\gamma_{el}\right\},
  \ee
  where
  \be\label{gamma}
  \gamma_{el}=\frac{e^2g_{+}^{2}-2ep_{\varphi}g_{-}}{1-e^2}+\frac{2aq_{+}}{e}
  \ee
  Hamiltonian written in this form is convenient for the further analysis of the corresponding Hamilton-Jacobi equations. However, in order to demonstrate the role and the structure
  of the emergent magnetic field with elliptic symmetry, one can recover the canonical form of the Hamiltonian with the explicit vector-potential:
  \begin{eqnarray}\label{HelA}
  && \mathcal{H}_{el}=\frac{e^2}{2a^2\left(1-e^2\eta^2
  \right)}\left\{(1-\eta^2)p_\eta^2+\frac{1-e^2\eta^2}{(1-e^2)(1-\eta^2)}\left(p_{\varphi}-A^{el}_{\varphi}\right)^2,
  +V_{el}(\eta)\right\},\\
  && A^{el}_{\varphi}=eg_-\frac{1+g_{el}^2-(\eta-g_{el})^2}{1-e^2\eta^2}, \quad \quad g_{el}=\frac{g_+ (1-e^2)}{2 e g_-}, \nonumber \\
  && V_{el}(\eta)=2 a q_-\eta+g_-^2 \frac{(1-e^2)(1-e^2\eta^2)-e^2(1+2 g_{el}\eta-\eta^2)^2}{(1-e^2)(1-\eta^2)(1-e^2\eta^2)}+\frac{e^2 g_+^2}{1-e^2}+\frac{2 a q_+}{e},\nonumber
  \end{eqnarray}
According to the general Hamilton-Jacobi
formalism, we look for a generating function in the following form:
\be\label{el_S_eta}
S_{el}=-Et+p_{\varphi}\varphi+S_{\eta},\qquad S_\eta=\int p_\eta
d\eta=\frac{1}{e\sqrt{1-e^2}} \int \sqrt{F}\frac{d\eta}{1-\eta^2}.
\ee
where
\bea
&F&=2a^2(1-e^2)(1-e^2\eta^2)(1-\eta^2)E-e^2(1-e^2\eta^2)p_{\varphi}^2-\gamma_{el}(1-e^2)(1-\eta^2)+\nonumber\\
&&+2(1-e^2)(p_{\varphi}g_+
-aq_-)\eta+2aq_-(1-e^2)\eta^3-g_-^2(1-e^2)\label{F}\eea
This system is a natural candidate for
the Landau problem on
the ellipsoid of revolution.
 The form of a trajectory in general case can be
rather complicated. However, we are looking for the specific values of the
$q_{1,2}$ and $g_{1,2}$ at which the integral in (\ref{el_S_eta})
qualitatively coincides with that of a free particle presented in
Eq.(\ref{S_el_eta}). It is easy to see that
this would be the case if the magnetic and electric charges of the dyons are related to each other by the following conditions:
\begin{itemize}
 \item $q_{-}=0$, $g_+ =0$.
\end{itemize}
 Equivalently, $q_1=q_2=q$, $g_1=-g_2=g$.
From the physical point of view here
we have  a magnetic dipole, or just a magnetic moment pointing
along the $z$-axis, ${\mathbf{m}}=(0,0,2ag)$. Of
cause, the specific condition of the same electric charges of the
dyons holds also for the case of zero charge, when one has just a
magnetic moment without electric field. Thus, one can speak about the reducible case of the Landau problem
on the ellipsoid of revolution. The corresponding vector-potential and potential term in the Hamiltonian from
Eq. (\ref{HelA}) are given by the following expressions:
\begin{eqnarray}\label{el_red}
 &&\tilde{A}^{el}_{\varphi}=2 e g \frac{1-\eta^2}{1-e^2 \eta^2}, \\
 && \tilde{V}_{el}(\eta)=4 g^2 \frac{1+e^2\left\{\left(1+e^2-\eta^2 \right)\eta^2-2\right\}}{(1-e^2)(1-\eta^2)(1-e^2\eta^2)}+\frac{4 aq}{e}.\nonumber
\end{eqnarray}
For this configuration the expression in (\ref{F}) simplifies to
\be
F=2a^2e(1-e^2)(1-e^2\eta^2)(1-\eta^2)E-e^3(1-e^2\eta^2)p_{\varphi}^2+
4(p_{\varphi}e^2g-aq(1-e^2))(1-\eta^2)-4g^2e(1-e^2)\ee
 Defining the new variable
$1-\eta^{2}=x$,  it can be transformed to the form of the partial
generating functions for the free particle on ellipsoid (See Eq.
(\ref{S_el_eta_x})), where the constants $b_{1,2}$ are now defined  by the
expressions \be\label{bmod}
b_{1}=\frac{2Ea^2e(1-e^2)^2-4aq(1-e^2)-e^{5}p_{\varphi}^{2}+4p_{\varphi}ge^2}{2Ea^2(1-e^2)e^3},\qquad
b_{2}=-\frac{e^2p_{\varphi}^{2}+4g^{2}}{2Ea^2e^2}. \ee
Thus,
 the motion of the charged particle on the ellipsoid of revolution with the
  magnetic dipole placed in the line connecting the foci of the ellipsoid and carrying
 equal electric charges at the poles is qualitatively  the same as for the free particle.
  The difference is only quantitative, stemming out from the different form of the constants $b_1$ and $b_2$.
  By this reason we refer it   as "reducible Landau problem on the ellipsoid of revolution" .
\subsubsection*{Action variables}
Since we deal with finite motion, we can construct the action variables for our  system\cite{arnold},
 \be\label{el_MICZ_av}
 I_{1}(p_{\varphi},E)=\frac{1}{2\pi}\oint p_{\eta}d\eta, \qquad
I_{2}=p_{\varphi}. \ee
After the substitution $x=\eta^2$ the first expression takes the form (the definition of the circle in $I_1$ , and
  details  of related calculations are completely similar to those presented in \cite{BNY_12})
   \be
   I_{1}(p_{\varphi},\eta)=
   \frac{a}{2\pi}\sqrt{\frac{E}{2}}\oint\sqrt{\frac{(x-a_+)(x-a_-)}{x(1-x)^2}}dx=
   \frac{a}{\pi}\sqrt{\frac{E}{2}}\int_{0}^{a_-}\sqrt{\frac{(x-a_+)(x-a_-)}{x(1-x)^2}}dx,
   \ee
where the values of the variable $x$ at the turnoff points are
   \be
   a_{\pm}=\frac{2+b_1\pm\sqrt{b_1^2-4b_2}}{2},
   \ee
where $b_{1,2}$ are defined by (\ref{bmod}).
One can see that $0<a_-<1$ and $a_+>1$. As $\eta<1$, we eventually get
   \be
   I_{1}(p_{\varphi},E)=aa_-\sqrt{\frac{a_+E}{2}}F_{1}\left(\frac{1}{2},1,-\frac{1}{2},2,a_-,\frac{a_-}{a_+}\right),
   \ee
 where $F_{1}\left(\frac{1}{2},1,-\frac{1}{2},2,a_-,\frac{a_-}{a_+}\right)$
   is the Appel hypergeometric function which is defined as follows:
   \be
   F_{1}\left(\alpha,\rho,\lambda,\alpha+\beta,ua,va\right)=a^{1-\alpha-\beta}\frac{\Gamma(\alpha+\beta)}
   {\Gamma(\alpha)\Gamma(\beta)}\int_{0}^{a}x^{\alpha-1}(a-x)^{\beta-1}(1-ux)^{-\rho}(1-vx)^{-\lambda}dx.
   \ee
\subsection*{The Landau problem on hyperboloid}
Let us now consider another two-dimensional system, which can be
obtained form the initial two-center MICZ-Kepler Hamiltonian.
Setting $\eta=const=\frac{1}{e}$ and $p_{\eta}=0$ ($e>1$) in Eq.
(\ref{Ham_MICZ}), we get the Hamiltonian of a charged particle moving
on the hyperboloid with two dyons placed on its foci:
\begin{eqnarray}
  \mathcal{H}_{hyp}=\frac{e^2}{2a^2\left(1-e^2\xi^2
  \right)}\left\{(1-\xi^2)p_\xi^2+\frac{1-e^2\xi^2}{(1-\xi^2)(1-e^2)}p_{\varphi}^2+
  \frac{g_{+}^{2}-2p_{\varphi}g_{-}\xi}{1-\xi^{2}}-2aq_{+}\xi+\gamma_{hyp}
  \right\}
  \end{eqnarray}
where the constant is
  \begin{eqnarray}
  \gamma_{hyp}=\frac{e^2g_{-}^{2}-2ep_{\varphi}g_{+}}{1-e^2}-\frac{2aq_{-}}{e}
  \end{eqnarray}
 As in the case of the ellipsoid let us explicitly  show  the vector-potential of the corresponding magnetic field by rewriting the Hamilton in the form similar to Eq. (\ref{HelA}).
  \begin{eqnarray}\label{HhypA}
   &&\mathcal{H}_{hyp}=\frac{e^2}{2a^2\left(1-e^2\xi^2
  \right)}\left\{(1-\xi^2)p_\xi^2+\frac{1-e^2\xi^2}{(1-\xi^2)(1-e^2)}\left(p_{\varphi}-A^{hyp}_{\varphi}\right)^2+V_{hyp}(\xi)
  \right\}, \\
  && A^{hyp}_{\varphi}=eg_+\frac{1+g_{hyp}^2-(\xi-g_{hyp})^2}{1-e^2\xi^2}, \quad \quad g_{hyp}=\frac{g_- (1-e^2)}{2 e g_+}, \nonumber \\
  && V_{hyp}(\xi)=-2 a q_+\eta+g_+^2 \frac{(1-e^2)(1-e^2\xi^2)-e^2(1+2 g_{hyp}\xi-\xi^2)^2}{(1-e^2)(1-\xi^2)(1-e^2\xi^2)}+\frac{e^2 g_-^2}{1-e^2}-\frac{2 a q_-}{e}.\nonumber
  \end{eqnarray}
In is worth mentioning that the Landau problem on ellipsoid and on hyperboloid can be transformed to each other via the following changes:
\begin{eqnarray}\label{trans}
 g_+\leftrightarrow g_-,\quad \quad q_+\leftrightarrow -q_-, \quad \quad \eta\leftrightarrow \xi.
\end{eqnarray}
The corresponding generating function reads
\be
S_{hyp}=-Et+p_{\varphi}\varphi+S_{\xi},
\qquad S_\xi=\int p_\xi
d\xi=\frac{1}{e\sqrt{1-e^2}} \int \sqrt{F}\frac{d\xi}{1-\xi^2}, \ee
where
\bea\label{Fh}
&F&=2a^2(1-e^2)(1-e^2\xi^2)(1-\xi^2)E-e^2(1-e^2\xi^2)p_{\varphi}^2-\gamma_{hyp}(1-e^2)(1-\xi^2)\nonumber\\
&&+2(1-e^2)(-p_{\varphi}g_-
-aq_+)\xi+2aq_-(1-e^2)\xi^3+g_+^2(1-e^2)\eea
As in the case of Landau problem on ellipsoid, here we are going to
find such values of the electric and magnetic charges of two dyons
placed at the foci of the hyperboloid that the emergent trajectories
of the particle qualitatively are the same as in the free case given by Eq.(\ref{sh}). It is
easy to see that this corresponds to the following configuration of the charges:
\begin{itemize}
 \item  $q_{+}=0$ and $g_{-}=0$.
\end{itemize}
Equivalently, $q_1=-q_2=q$, $g_1=g_2=g$.
This  means that
the electric and magnetic fields correspond to the electric dipole pointing
along the $z$-axis,
with  $\mathbf{d}=(0,0,2aq)$, whose poles are carrying the magnetic charges $g=g_1=g_2$.
In this reducible case of the Landau problem on the hyperboloid of revolution, the vector-potential and the potential term in the Hamiltonian form Eq. (\ref{HhypA}) take the
following forms:
\begin{eqnarray}\label{hyp_red}
 &&\tilde{A}^{hyp}_{\varphi}=2 e g  \frac{1-\xi^2}{1-e^2 \xi^2}, \\
 && \tilde{V}_{hyp}(\xi)=4 g^2 \frac{1+e^2\left\{\left(1+e^2-\xi^2 \right)\xi^2-2\right\}}{(1-e^2)(1-\eta^2)(1-e^2\eta^2)}-\frac{4 aq}{e}.\nonumber
\end{eqnarray}
Then the expression (\ref{Fh}) simplifies to \be
F=2a^2e(1-e^2)(1-e^2\xi^2)(1-\xi^2)E-e^3(1-e^2\xi^2)p_{\varphi}^2+4(p_{\varphi}e^2g-
aq(1-e^2))(1-\xi^2)-4g^2e(1-e^2).\ee After the appropriate change of
the variable, $1-\xi^{2}=x$, the corresponding generating function
takes the form (\ref{bmod}), with \be\label{bmod}
b_{1}=\frac{2Ea^2e(1-e^2)^2-4aq(1-e^2)-e^{5}p_{\varphi}^{2}+4p_{\varphi}ge^2}{2Ea^2(1-e^2)e^3},\qquad
b_{2}=-\frac{e^2p_{\varphi}^{2}+4g^{2}}{2Ea^2e^2}. \ee In full
analogy with the previous case of ellipsoid,
 the character of the motion on hyperboloid under the presence of
  the electric dipole with equal magnetic charges at the poles
 is qualitatively the same as a free motion on hyperboloid. Thus, the constructed system can be called "reducible Landau problem on the hyperboloid of revolution".
In case of hyperboloid it is impossible to construct action-angle
variables, because the motion is infinite.

\subsection*{Landau problem on paraboloid}
The two-center MICZ-Kepler Hamiltonian can be used
for the constructing of the Landau problem on paraboloid.
  To do so, one should consider a limiting case of two-center MICZ-Kepler Hamiltonian, written down in parabolic coordinates (\ref{Ham_par_ZS}) where one of the dyons
  is placed at infinity .
It  can be used, particularly, for the description of the Zeeman-Stark effect in the presence of Dirac monopole \cite{BO}.
Let us now set  $\eta=const=\frac{p}{2}$ and $p_{\eta}=0$. The
corresponding Hamiltonian describes the particle (with unit electric charge) moving on
the surface of the paraboloid with the electric charge $q$ and magnetic charge $g$ placed in its focus and with
the parallel constant uniform electric and magnetic fields ${\cal E}$ and  $B$ directed along the $z$-axis:
\begin{eqnarray} \label{Ham2}
 \mathcal{H}_{par}=\frac{1}{p+2\xi}\left\{4\xi p_{\xi}^{2}+
\frac{(p_\varphi+g)^2}{\xi}+3gB\xi-
 {\cal E} \xi^2+\frac{B^2 \xi^3}{4}+\gamma_{par}\right\}-\frac{B p_{\varphi}}{2},
 \end{eqnarray}
where
 \begin{eqnarray}
 \gamma_p=4q+
  \frac{2(p_{\varphi}-g)^2}{p}+\frac{p}{32}\left(B\left(p^2B-48g\right)+8p{\cal E}\right)
 \end{eqnarray}
 Us usual, let us rewrite the Hamiltonian in such a way that the explicit form of the contribution from the magnetic field with parabolic symmetry is visible.
 \begin{eqnarray}\label{HparA}
  &&\mathcal{H}_{par}=\frac{1}{p+2\xi}\left\{4\xi p_{\xi}^{2}+\frac{p+2 \xi}{p \xi}\left(p_\varphi-A^{par}_{\varphi}\right)^2+V_{par}(\xi)\right\}, \\
  && A^{par}_{\varphi}=\frac{p B}{2}\frac{g_{par}^2-\frac{2 g}{B}-(\xi-g_{par})^2}{p+2\xi}, \quad \quad g_{par}=\frac{2 g}{pB}-\frac{p}{4}, \nonumber \\
  && V_{par}(\xi)=\frac{g^2}{\xi}+3 g B \xi- {\cal E} \xi^2+\frac{B^2 \xi^3}{4}-\frac{p B^2}{4}\frac{\left( \frac{2 g}{B}+2 g_{par}\xi-\xi^2\right)^2}{\xi (p+2 \xi)}+
  +4 q+\frac{2 g^2}{p}+\frac{p}{32}\left( B (p^2 B-48 g)+8 p {\cal E} \right). \nonumber
 \end{eqnarray}
Its  generating function reads
\begin{eqnarray}
 S_{par}=-\left(E+\frac{1}{2}p_{\varphi}B\right)t+p_{\varphi}\varphi+S_{\xi},
\end{eqnarray}
where \be S_{\xi}=\int\sqrt{-B^2\xi^4+4{\cal
E}\xi^3+8(-gB+E+\frac{1}{2} p_\varphi
B)\xi^2+4(-\gamma_p+pE+\frac{1}{2}pp_\varphi
B)\xi-4(p_\varphi+g)^2}\frac{d\xi}{4\xi} \ee
This generating function takes the same form as for
the case of free particle given by Eq. (\ref{S_par_xi}), if we choose ${\cal E}=0$,
$B=0$, which leads  to the following values of the constants:
\be
b_{1}=\frac{p}{2}-\frac{2qp+(p_\varphi-g)^2}{Ep},\;\;b_{2}=-\frac{\left(p_{\varphi}+g\right)^2}{2E},
 \ee
 where $q$ and $g$ are the electric and magnetic charges  of the dyon placed on its focus.
Since the form   of the generating function  in this case is the same as in the case
  of the free particle on paraboloid, one can conclude that the
  inclusion of the dyon in the focus does not change the character of motion qualitatively.
  Only quantitative changes are expected.
The vector-potential and the potential term of the Hamiltonian in this particular case take the following form:
\begin{eqnarray}
 &&\tilde{A}_{\varphi}^{par}=g \frac{\xi-p/2}{\xi+p/2}, \\
 && \tilde{V}_{par}(\xi)=g^2\left\{\frac{1}{\xi}+\frac{2}{2}\left(1-\frac{(\xi-p/2)^2}{\xi+p/2}\right)\right\}+4q. \nonumber
\end{eqnarray}
Therefore, we call the constructed system "reducible Landau problem on the paraboloid of revolution".
\section{Summary and Discussion }

In this paper we considered the problem of formulating the Landau
problem on two-dimensional ellipsoid, hyperboloid and
paraboloid of revolution. We succeeded in resolving this problem thanks to the
integrable two-center MICZ-Kepler system (generalization of
two-center Kepler system with attractive centers carrying both
electric and magnetic charges) introduced in Refs.\cite{KNO}.
 As this  system
admits the separation of variables in elliptic coordinates,
the cases of the particle  on
ellipsoid and hyperboloid were  obtained just by the
restriction of the whole three-dimensional systems to one of the
coordinate surfaces which are coaxial ellipsoids and
hyperboloids of revolution for the elliptic coordinate system. The parabolic case
is also obtained in a straightforward way by considering the limiting case
of the two-center MICZ-Kepler Hamiltonian where one of the foci is
sent to infinity.  Then, we found  the constrains
for the system parameters upon which the classical trajectories of
the particle share the same qualitative features with the free motion
on the corresponding surfaces. We found that for the case of
ellipsoid the electric charges of two dyons should be the same,
while magnetic ones should be of the same magnitude but with opposite
signs. Thus, this situation is equivalent to the particle moving in
the filed of a magnetic moment, which can be implemented
experimentally. The hyperbolic case is the opposite: electric dipole and
two identical magnetic charges.

 It is worth mentioning that the two-center MICZ-Kepler system, besides its integrability,
 possess one more important feature. It admits   ${\cal N}=4$ supersymmetric  extension, provided the magnetic and electric charges of the dyons
 satisfy the condition
 (the corresponding issues as well as the variant of the supersymmetric MICZ-Kepler system on $S^3$ have been considered in \cite{KNO,denef,NO,BKO}):
 \begin{equation}\label{DSZ}
 {q_1}{g_2}={q_2}{g_1}.
 \end{equation}
Thus, the contraction of the corresponding supersymmetric systems to
the corresponding two-dimensional surfaces will lead to a
supersymmetric Landau problems on ellipsoid, hyperboloid and
paraboloid, however, due to the restriction on the values of
electric and magnetic charges found in this work, the condition
(\ref{DSZ}) requires the vanishing of the electric charges of the
background dyons. Consideration of this problem should be subject of
a separate study and will
 be considered elsewhere\\

{\large Acknowledgements}
 This work was done within the  joint grant of the Armenian
State Committee of Science  and Belorussian Foundation for Basic
Research 11AB-001 and  by  Volkswagen Stiftung under contract No. 86 260.
Aremnian participants also acknowledge the partial financial support form the grants of the State Committee of Science of Armenia,
No. 13-1C114 and No. 13-1F343.

\end{document}